# Learning-Based Control Compensation for Multi-Axis Gimbal Systems Using Inverse and Forward Dynamics

**DAMLA LEBLEBICIOGLU**
Bilkent University, Ankara, Turkey

**ÖZGÜR ATESOGLU**
ROKETSAN Missiles Inc., Ankara, Turkey

**ANIL E. DERINOZ**
ROKETSAN Missiles Inc., Ankara, Turkey

**MELİH CAKMAKCI**
Bilkent University, Ankara, Turkey

Author's addresses: Damla Leblebicioğlu and Melih Çakmakcı are with the Department of Mechanical Engineering, Bilkent University, Ankara, 06800 Turkey, E-mail: (d.leblebicioglu, melihc@bilkent.edu.tr). Özgür Ateşoğlu and Anıl E. Derinöz are with the ROKETSAN Missiles Inc., Ankara, 06780 Turkey, E-mail: (ozgur.atesoglu, anil.derinoz@roketsan.com.tr).
*(Corresponding author: Melih Cakmakci)*

**Unmanned aerospace vehicles usually carry sensors (i.e., electro-optical and/or infrared imaging cameras) as their primary payload. These sensors are used for image processing, target tracking, surveillance, mapping, and providing high-resolution imagery for environmental surveys. It is crucial to obtain a steady image in all these applications. This is typically accomplished by using multi-axis gimbal systems. This paper concentrates on the modeling and control of a multi-axis gimbal system. A novel and fully outlined procedure is proposed to derive the nonlinear and highly coupled Equations of Motion of the two-axis gimbal system. Different from the existing literature, Forward Dynamics of the two-axis gimbal system is modeled using multi-body dynamics modeling techniques. In addition to the Forward Dynamics model, the Inverse Dynamics model is developed to estimate the complementary torques associated with the state and mechanism-dependent, complex disturbances acting on the system. A disturbance compensator based on multilayer perceptron (MLP) structure is implemented to cope with external and internal disturbances and parameter uncertainties through the torque input channel. Our initial simulations and experimental work show that the new NN (neural network)-based controller is performs better in the full operational range without requiring any tuning or adjustment when compared with well-known controllers such as cascaded PID, ADRC (Active Disturbance Rejection Control), Inverse Dynamics based controllers.**

## I. INTRODUCTION

In past few decades, civil and military applications of unmanned aerospace vehicles have been steadily increasing. Unmanned aerospace vehicles (such as rockets, drones, and satellites) usually carry sensors as their primary payload. Hence, electro-optical and/or infrared imaging cameras are used for image processing, target tracking, surveillance, mapping, and providing high-resolution imagery for environmental surveys [1].

A common use case scenario is given in Fig. 1 for surveillance of cargo ships. As expected, it is crucial to obtain a steady image in such applications, which is typically accomplished by using multi-axis gimbal systems. Gimbal Systems ensure the Line-of-Sight (LOS) and Field of View (FOV) of the camera payload stays stable and stationary when oriented towards a desired target or point of interest, [2]. The operation should also be robust to disturbances originated by the motion of the host vehicle. Furthermore, in coordination with the tracking algorithm, the gimbal system steers the LOS vector onto the target and provides information to estimate its relative angular position [1], [3]. With that information, the vehicle is guided along the desired trajectory or onto the target.

The two-axis gimbal system studied in our work carries an optical sensor payload as shown in Fig. 2. A two-axis gimbal system is a two-degree of freedom (2-DoF) chain mechanism in which revolute joints connect the inner gimbal to the outer gimbal and the outer gimbal to the base platform. The differential equations that define the motion of this multi-body system are nonlinear, axis dynamics are highly coupled and therefore, difficult to model with high fidelity for control development studies.

So far, the research that focuses on mathematical modeling of gimbal platforms commonly uses various assumptions to simplify the Equations of Motion (EoMs). Mathematical models of gimbals can be obtained by following either the Newton-Euler formulation or the Lagrangian approach [4], [6]. In [4], it is assumed that rotation axes of inner and outer gimbals coincide at a single point and the center of gravity (CoG) of the whole mechanism is located at that same pivot point. In [6], it is considered that there is an offset between the center of gravity of the whole system and that pivot point, but it is still assumed that rotation axes of inner and outer gimbals coincide. Some research papers present plant dynamics using only mass moment of inertia terms [10]-[13], [23]. In [7], [8] and [25], researchers have simplified their gimbal models by neglecting dynamical mass unbalance (i.e. neglecting that the inertia tensors of the inner and outer gimbals contain off-diagonal terms), which reduced the cross-coupling effects. In [9] and [22], the impact of dynamical mass unbalance is included, but the center of gravity offsets, rotation axes misalignments and disturbance forces/moments are not explicitly modeled.

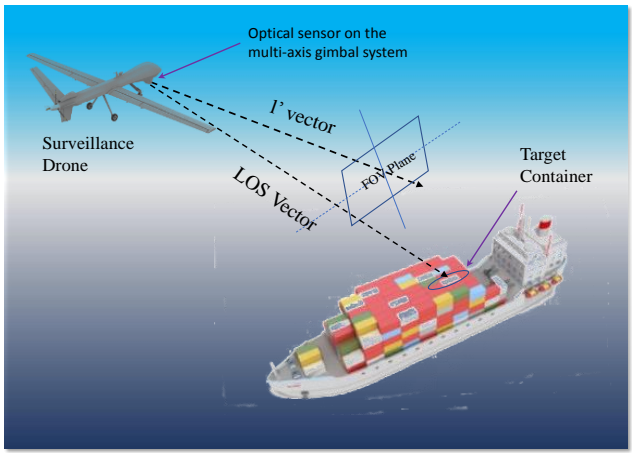

Fig. 1. Field of View (FOV) is the angular area that a camera can observe at any given time when the gimbal system is stationary. Aim of the gimbal system is to hold the image of the target at the center of the FOV plane by aligning LOS vector with the $l'$ vector and minimizing the error ($\varepsilon$) between them while the vehicle is moving

In [14], [15], the cross-coupling terms, friction forces/moments on the joints and other sources of model uncertainties, external disturbances are counted together to generate collective disturbance torques, acting on the gimbal joints. The two-axis gimbal platform model used in [16] and [17] includes joint reaction moments on azimuth and elevation gimbals. However, the inertia matrices are composed only of diagonal terms and no information on the center of gravity offsets or, rotation axes misalignments are provided. The two-axis gimbal model proposed in [18] includes mass unbalance as a disturbance torque for both inner and outer gimbals, but this does not explicitly mention the dynamical mass unbalance generated by the off-diagonal terms of the moment of inertia matrices of the inner and outer gimbals. [18], rather defines joint reaction forces as unbalance torques. Considering the related literature and references, [1], [19] and [20] present the most sophisticated and fundamental models, regarding their contribution on the design principles of multi-axis gimbal systems, derivation of detailed kinematic and dynamic equations, expressions on cross-coupling effects, reaction moments and disturbance torques. However, those articles do not consider the misalignments between rotation axes of gimbals, the offsets between the center of gravity of the gimbals and the rotation axes, the implications of having a 3D system dynamics on the mathematical model (the moment of inertia matrices of the gimbals that contain off-diagonal terms), the disturbances on the joints arising from friction or some other restraining elements of the whole gimbal system (such as cables, wires, pipes) and the reaction forces and moments on the connection joints of the gimbal system in detail.

A two-axis gimbal is a MIMO (multiple-input and multiple-output) complex nonlinear system that is susceptible to the influence of unknown frictional torques, external disturbances and system uncertainties. Thus, various control strategies have been developed in literature to satisfy the accurate positioning requirements during target tracking. Control of a two-axis gimbal system is generally implemented using a three-loop structure. The outer loop is the tracking loop and it is the position controller. Tracking loop minimizes the error between the LOS vector and the l' vector (Fig. 1). The middle loop is the stabilization loop controlling the rate. Stabilization loop eliminates base disturbances after the host vehicle locks-on to the target. The inner loop controls the current produced by the motor. Traditional and straightforward control approaches such as proportional-integral-derivative (PID) controller, cascaded PID controller or proportional-integral (PI) controller [4], [6] are used in tracking and stabilization loops, over the past decade. When nonlinear frictional effects are significant, classical controllers are insufficient for compensation due to their limited capability, [2], [7], [8], [15], [18], [22], and [23]. In [24], $H_\infty$, Linear Quadratic Gaussian/Loop Transfer Recovery (LQG/LTR) and μ-synthesis controllers are designed for the stabilization loop by using the gimbal model proposed in [20]. Since dynamical mass unbalance and cross-coupling terms are disregarded in [20], the controller design is performed with a simpler mathematical model. In [13], the LuGre model is used to represent the nonlinear friction behavior present in an inertially stabilized platform (ISP). Friction model parameters are identified and a backstepping integral adaptive compensator is designed to compensate for this disturbance. In [5], [7] and [23], Active Disturbance Rejection Control (ADRC) approach, has been implemented as an alternative to replace and eliminate the limitations of conventional PID control. ADRC theory has been one of the popular and useful methods in controller design for gimbal systems since its proposal by J. Han in [29]. Even though ADRC implementation with linear extended state observer (ESO) has less parameters to tune compared to nonlinear ESO design, two different ESOs will be designed for a two-axis gimbal system. SISO (single-input and single-output) ADRC controller is designed for each channel.

Neural network-based control schemes [8], [11], [25] are another outstanding trend in the adaptive gimbal control field. Radial Basis Function Neural Network (RBFNN) is used with SMC in [8] and with state feedback control in [25]. A neural network based adaptive technique is employed for dead-zone effect compensation caused by the limitations of actuators in [11]. Neural Networks (NNs) are effective tools for representing complex behaviors, possibly originating from parameter uncertainties, unmodeled dynamics and disturbance effects in nonlinear systems, [26].

In this paper, we use a neural network structure for supplementary compensation of external and internal disturbances and parameter uncertainties, through torque input channel. In addition to the nonlinear friction torque in the revolute joints, there are internal disturbances that are caused by plant parameter uncertainties. These mainly refer to disturbances because of CoG and rotation axis-offsets and off-diagonal terms in the mass moment of inertia matrices. Furthermore, the electrical cables of the motors, the encoders/resolvers, the gyroscope and the cooling pipes of the sensor system generate additional, gimbal pose and state dependent, time varying disturbances. Hence, it is very hard to accurately represent the disturbances in all operation regimes with a single model, as also observed in [30].

The primary contributions of this study can be listed as follows: (i) the gimbals are modeled as separate rigid bodies that can make 6-DoF yet constrained motion. Besides the customary rotational motion of the gimbals, the translational motion is also modeled, the offsets between the center of gravity of the gimbals and the rotation axes, the off-diagonal terms of the moment of inertia matrices, the disturbances on the joints and the reaction forces and moments on the connection joints of the gimbal system are all taken into account, (ii) Using this novel-model formulation, an "inverse" dynamics model is also developed to estimate the complementary torque values to compensate the gimbal pose dependent, time varying disturbances, (iii) Using data model manipulated using the inverse dynamics calculations, a multilayer perceptron (MLP) NN structure based disturbance compensator is implemented to cope with external and internal disturbances and parameter uncertainties through torque input channel. This approach distinguishes itself as a new compensator method that can be applied (iv) collecting and training data for the MLP and (v) calculating disturbance torque in the system is developed by using ideal inverse dynamics and forward dynamics (real plant) together in a successive manner.

The rest of the article is structured as follows: In Chapter II, an extensive mathematical model of a two-axis gimbal system is presented. Chapter III, introduces the gimbal test set-up, experimental data collection, and hardware implementation together with disturbance torque identification. The construction of the MLP structure and the performance is presented by simulations and experimental data in Chapter IV. In Chapter V, the results, and future work are discussed.

## II. MATHEMATICAL MODEL OF THE SYSTEM

In this section, the mathematical model of a typical two-axis gimbal system, shown in Fig. 2, is presented. Ateşoğlu et al. initially presented the modeling approach pursued in this article in [28]. In this study, their initial mathematical model is re-visited and extended further with the development of Forward and Inverse Gimbal Dynamics. The Forward Dynamics model calculates the motion of the gimbal using the driving torques, whereas the Inverse Dynamics model reverses the EOMs of the forward model and calculates the necessary torque required to perform a pre-defined motion.

Fig. 2 shows the two-axis gimbal platform and its payload studied in this work. The mechanical structure of the two-axis gimbal system is composed of three interconnected rigid bodies, the inner ring, the outer ring and the base platform. The primary task of the two-axis gimbal system is to rotate the payload, i.e., the camera mounted on the inner ring, with respect to the base platform. For that purpose, the mechanical design is conducted in such a way that the outer ring performs the azimuth and the inner ring performs the elevation motion.

Hence, the outer and inner rings are defined as the yaw and pitch gimbals, respectively.

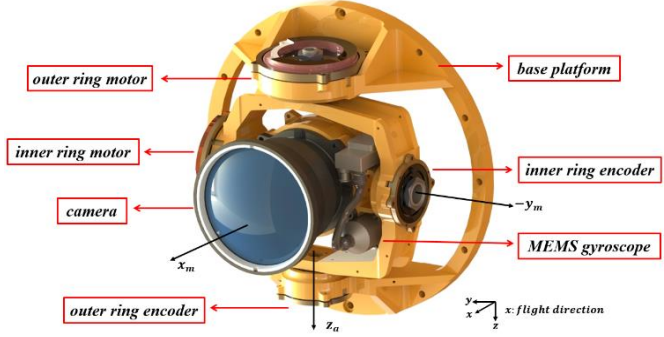

Fig. 2. Two-axis gimbal system carrying the camera

Two brushless direct current (DC) motors actuate the inner and outer rings of the two-axis gimbal system. Angular positions are measured by two absolute encoders mounted within the revolute joints, which are joining the yaw gimbal to the base platform and joining the pitch gimbal to the yaw gimbal, respectively. Thus, they measure the relative angular position of the yaw gimbal with respect to the base platform and the relative angular position of the pitch gimbal with respect to the yaw gimbal. Furthermore, a micro-electromechanical system (MEMS) gyroscope is installed on the pitch gimbal to measure the angular velocity of the pitch gimbal with respect to a chosen inertial reference frame.

The rotation between the base platform reference frame $F_b$ (with axes $x_b, y_b, z_b$) and the pitch gimbal (with the camera and the MEMS gyroscope mounted on) reference frame $F_m$ (with axes $x_m, y_m, z_m$) is defined with a 3-2 (yaw-pitch) rotation sequence. Here, two consecutive rotations should be conducted. First rotation is around the z-axis of the base platform reference frame with angle ($\psi_a$), the second rotation is around the y-axis of the rotated reference frame, i.e., the outer gimbal reference frame $F_a$ (with axes $x_a, y_a, z_a$), with angle ($\theta_m$). Kinematic Equations for the Rotational Motion of the two-axis gimbal are given in Appendix A.

### A. KINEMATIC EQUATIONS FOR THE TRANSLATIONAL MOTION OF THE TWO-AXIS GIMBAL

In a kinematical chain of multi-body system, the inertial properties of each bodies are considered separately [31]. In Figs. 3 and 4, the definition of the position vectors, which are used in the translational kinematic equations of motion of the two-axis gimbal system are shown. Ideally, the CoG of the yaw gimbal is on the rotation axis of the outer ring and the CoG of the pitch gimbal is on the rotation axis of the inner ring. Also, the pivot point is the intersection point of the rotation axis of the outer ring and the rotation axis of the inner ring.

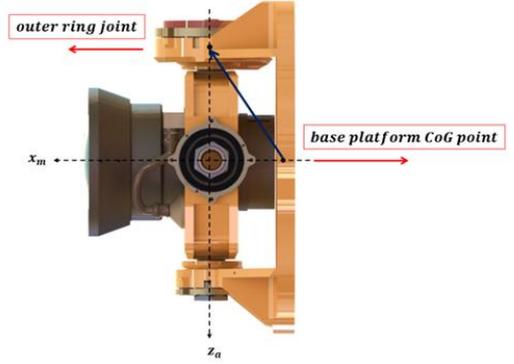

Fig. 3. The position vector from the Center of Gravity (CoG) of the base platform to the outer ring revolute joint

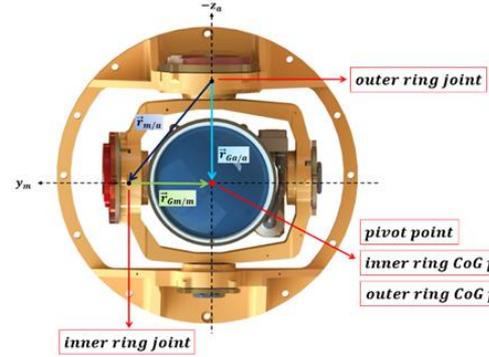

Fig. 4. The Position vector from the outer ring revolute joint to the CoG of the yaw gimbal, $\vec{r}_{Ga/a}$, the position vector from the inner ring revolute joint to the CoG of the pitch gimbal, $\vec{r}_{Gm/m}$, and the position vector from the outer ring revolute joint to the inner ring revolute joint, $\vec{r}_{m/a}$

Due to assembling and production errors, or physical disturbance during the flight, the CoG position vectors change from their intended values (i.e., Fig. 4). In Fig. 5, the offset between rotation axes of inner and outer gimbals and some cases of displaced CoG points of the inner and outer gimbals from rotation axes are shown.

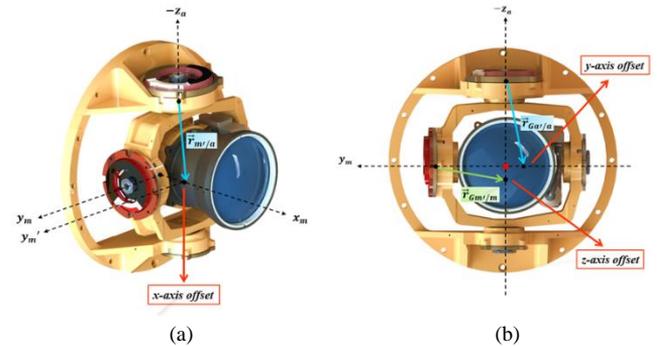

Fig. 5. The position vectors $\vec{r}_{m'/a}$ (a), $\vec{r}_{Ga'/a}$ and $\vec{r}_{Gm'/m}$ (b) for disturbed (non-ideal) cases

The position vector defining the CoG of the yaw gimbal with respect to a chosen inertial frame ($\vec{r}_{Ga/o}$) is the sum of three position vectors: the position vector of the CoG of the base platform with respect to a chosen inertial frame ($\vec{r}_{b/o}$), the vector defining the position of the outer ring revolute joint with respect to the CoG of the base platform ($\vec{r}_{a/b}$) and the position vector of the CoG of the yaw gimbal with respect to the outer ring revolute joint ($\vec{r}_{Ga/a}$).

$$\vec{r}_{Ga/o} = \vec{r}_{Ga/a} + \vec{r}_{a/b} + \vec{r}_{b/o} \quad (1)$$

In column matrix notation, we can write the same equation as shown below. Here, the components of the vectors are expressed at outer gimbal frame, $F_a$:

$$\bar{r}_{Ga/o}^{(a)} = \bar{r}_{Ga/a}^{(a)} + \hat{R}_3(-\psi_a)\bar{r}_{a/b}^{(b)} + \hat{R}_3(-\psi_a)\bar{r}_{b/o}^{(b)} \quad (2)$$

By taking the time derivative of (1) twice with respect to the inertial reference frame and assuming that the base platform and the yaw gimbal are rigid bodies, we end up with the expression for the translational acceleration of the CoG of the yaw gimbal, with respect to the inertial reference frame, $F_o$.

$$\bar{a}_{Ga/o}^{(a)} = \ddot{\psi}_a \tilde{u}_3 \bar{r}_{Ga/a}^{(a)} + \bar{D}a_{Ga}^{(a)} \quad (3)$$

Here, note that, since the base platform and the yaw gimbal are assumed as rigid bodies, first and second time derivatives of $\bar{r}_{a/b}^{(b)}$ and $\bar{r}_{Ga/a}^{(a)}$ are zero vectors. On the other hand, the translational acceleration of the base platform with respect to the inertial reference frame is defined as $\bar{a}_{b/o}^{(b)} = [a_x \; a_y \; a_z]^T$. $\bar{D}a_{Ga}^{(a)}$ in (3) is used to shorten the expression of $\bar{a}_{Ga/o}^{(a)}$, explicitly given in App. A, (A9). Similarly, the position vector defining the CoG of the pitch gimbal with respect to a chosen inertial reference frame, $\vec{r}_{Gm/o}$, is the sum of four position vectors.

$$\vec{r}_{Gm/o} = \vec{r}_{Gm/m} + \vec{r}_{m/a} + \vec{r}_{a/b} + \vec{r}_{b/o} \quad (4)$$

In (4), $\vec{r}_{Gm/m}$ is the position of the CoG of the pitch gimbal with respect to the inner ring revolute joint and $\vec{r}_{m/a}$ is the position of the inner ring revolute joint with respect to the outer ring revolute joint. In column matrix notation, we can write the same equation as shown below. Here, the components of the vectors are expressed at $F_m$:

$$\bar{r}_{Gm/o}^{(m)} = \bar{r}_{Gm/m}^{(m)} + \hat{R}_2(-\theta_m)\bar{r}_{m/a}^{(a)} + \hat{R}_2(-\theta_m)\hat{R}_3(-\psi_a)\bar{r}_{a/b}^{(b)} + \hat{R}_2(-\theta_m)\hat{R}_3(-\psi_a)\bar{r}_{b/o}^{(b)} \quad (5)$$

By taking the time derivative of both sides of (5) twice with respect to the inertial reference frame and assuming that the yaw and pitch gimbals are rigid bodies, we end up with the expression for the translational acceleration of the CoG of the pitch gimbal, with respect to the inertial reference frame.

$$\bar{a}_{Gm/o}^{(m)} = \ddot{\psi}_a [\hat{R}_2(-\theta_m)\tilde{u}_3 \bar{r}_{m/a}^{(a)} + \widetilde{D_1 \bar{a}_{m/o}^{(m)}} \bar{r}_{Gm/m}^{(m)}] + \ddot{\theta}_m \tilde{u}_2 \bar{r}_{Gm/m}^{(m)} + \bar{D}a_{Gm}^{(m)} \quad (6)$$

Here note that, since the yaw and pitch gimbals are assumed as rigid bodies, first and second time derivatives of $\bar{r}_{m/a}^{(a)}$ and $\bar{r}_{Gm/m}^{(m)}$ are $\bar{0}$. Also, $\bar{D}a_{Gm}^{(m)}$ is used to shorten the expression of $\bar{a}_{Gm/o}^{(m)}$, explicitly given in App. A, (A10). Introduction of distance vectors, $\vec{r}_{a/b}, \vec{r}_{m/a}, \vec{r}_{Ga/a}$ and $\vec{r}_{Gm/m}$, in the mathematical model is an implication of existence of static mass unbalance.

## B. DYNAMICAL EQUATIONS FOR THE TRANSLATIONAL MOTION OF THE TWO-AXIS GIMBAL

Since the components considered are rigid bodies in 3D motion, $\hat{J}_a$, and the moment of inertia matrix of the pitch gimbal, $\hat{J}_m$ are both $3x3$ matrices with cross-axis diagonal terms, implying that dynamical mass unbalance is not neglected as shown in (7).

$$\hat{J}_a = \begin{bmatrix} J_{axx} & J_{axy} & J_{axz} \\ J_{ayx} & J_{ayy} & J_{ayz} \\ J_{azx} & J_{azy} & J_{azz} \end{bmatrix}, \hat{J}_m = \begin{bmatrix} J_{mxx} & J_{mxy} & J_{mxz} \\ J_{myx} & J_{myy} & J_{myz} \\ J_{mzx} & J_{mzy} & J_{mzz} \end{bmatrix} \quad (7)$$

Following the Newton-Euler approach, the mathematical model for the translational and rotational dynamics of the yaw gimbal can be derived as shown in (8) and (9).

$$m_a(\bar{a}_{Ga/o}^{(a)} - \bar{g}_a) = \bar{F}_{a/m}^{(a)} + \bar{F}_{a/b}^{(a)} \quad (8)$$

$$\hat{J}_a \bar{\alpha}_{a/o}^{(a)} + \tilde{\omega}_{a/o}^{(a)} \hat{J}_a \bar{w}_{a/o}^{(a)} = \bar{M}_{a/m}^{(a)} + \bar{M}_{a/b}^{(a)} + \tilde{r}_{a/m}^{(a)} \bar{F}_{a/m}^{(a)} + \tilde{r}_{a/b}^{(a)} \bar{F}_{a/b}^{(a)} \quad (9)$$

In (8), $\bar{g}_a$ is the gravitational acceleration vector represented at the yaw gimbal reference frame. Also, $\bar{r}_{a/m}^{(a)} = -\bar{r}_{m/a}^{(a)}$, $\bar{r}_{a/b}^{(a)} = \hat{R}_3(-\psi_a)\bar{r}_{a/b}^{(b)}$ in (9). Furthermore, in (9), $\bar{F}_{a/m}^{(a)} = [F_{amx} \; F_{amy} \; F_{amz}]^T$ is the reaction force on the revolute joint in between the yaw and pitch gimbals, and $\bar{F}_{a/b}^{(a)} = [F_{abx} \; F_{aby} \; F_{abz}]^T$ is the reaction force on the revolute joint in between the yaw gimbal and base platform. $\bar{M}_{a/m}^{(a)} = [M_{amx} \; M_{amy} \; M_{amz}]^T$ is the total moment on the revolute joint in between the yaw and pitch gimbals, and $\bar{M}_{a/b}^{(a)} = [M_{abx} \; M_{aby} \; M_{abz}]^T$ is the total moment on the revolute joint in between the yaw gimbal and base platform.

In $\bar{M}_{a/b}^{(a)}$, $M_{abz} = T_a - T_{fra}$, where $T_a$ is the driving torque applied on the yaw gimbal by the brushless DC

motor installed in between the yaw gimbal and the base platform, and $T_{fra}$ is the disturbance torque on the revolute joint in between the outer ring and the base platform. Similarly, in $\bar{M}_{a/m}^{(a)}$, $M_{amy} = T_e - T_{frm}$, where $T_e$ is the driving torque applied on the pitch gimbal by the brushless DC motor installed in between the pitch gimbal and the yaw gimbal, and $T_{frm}$ is the disturbance torque on the revolute joint between the inner and outer rings of the two-axis gimbal system.

The mathematical model for the translational and rotational dynamics of the pitch gimbal is also derived by applying the Newton-Euler approach:

$$m_m(\bar{a}_{Gm/o}^{(m)} - \bar{g}_m) = \bar{F}_{m/a}^{(m)} \qquad (10)$$

$$\hat{J}_m \bar{\alpha}_{m/o}^{(m)} + \tilde{\omega}_{m/o}^{(m)} \hat{J}_m \bar{\omega}_{m/o}^{(m)} = \bar{M}_{m/a}^{(m)} + \tilde{r}_{m/a}^{(m)} \bar{F}_{m/a}^{(m)} \qquad (11)$$

In (10), $\bar{g}_m$ is the gravitational acceleration vector represented at the pitch gimbal reference frame. Also, $\bar{r}_{m/a}^{(m)} = \hat{R}_2(-\theta_m)\bar{r}_{m/a}^{(a)}$, $\bar{F}_{m/a}^{(m)} = -\hat{R}_2(-\theta_m)\bar{F}_{a/m}^{(a)}$, $\bar{M}_{m/a}^{(m)} = -\hat{R}_2(-\theta_m)\bar{M}_{a/m}^{(a)}$, in (11). By rearranging (8)-(11), the resultant dynamical equations for the multi-body dynamics of the two-axis gimbal system is composed in matrix equation form as shown in (12). The matrices, $\hat{F}$, $\hat{R}$, $\hat{D}$ and $\hat{G}$ (in Eqn. 12) are given in the open form in Appendix A, (A11)-(A14).

$$\hat{F}_{12x2} \begin{bmatrix} \ddot{\psi}_a \\ \ddot{\theta}_m \end{bmatrix}_{2x1} + \hat{R}_{12x10} \begin{bmatrix} \bar{F}_{a/m}^{(a)} \\ \bar{F}_{a/b}^{(a)} \\ M_{amx} \\ M_{amz} \\ M_{abx} \\ M_{aby} \end{bmatrix}_{10x1} = \hat{D}_{12x1} + \hat{G}_{12x2} \begin{bmatrix} T_a \\ T_e \end{bmatrix}_{2x1} \qquad (12)$$

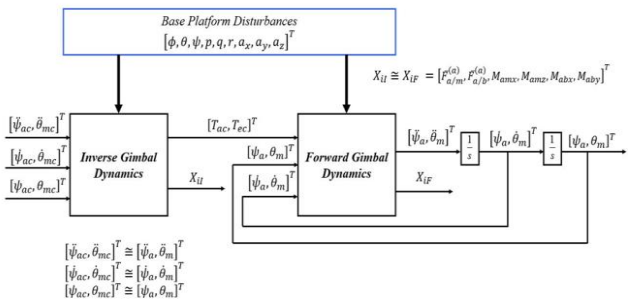

Fig. 6. Block diagram shows complementary nature of the *Inverse* and *Forward Dynamics* of the two-axis gimbal system

Eqn. 12 is the compact form of the nonlinear, and coupled, differential equations that represent the *Forward Dynamics* of the two-axis gimbal system. With the forward dynamics equations, $\ddot{\psi}_a$, $\ddot{\theta}_m$, can be calculated using the driving, or the control torques, $T_a$, $T_e$, the relative angular accelerations of the yaw and pitch gimbals. Also, thanks to the Newton-Euler approach, the reaction forces ($\bar{F}_{a/m}^{(a)}$, $\bar{F}_{a/b}^{(a)}$) and the constraining components of reaction moments ($M_{amx}$, $M_{amz}$, $M_{abx}$, $M_{aby}$) can also be calculated here as by-products. Using Eqn. 12, an equation representing the *Inverse Dynamics* of the two-axis gimbal system can also be given as shown in (13).

$$[-\hat{G}_{12x2} \vdots \hat{R}_{12x10}] \begin{bmatrix} T_{ac} \\ T_{ec} \\ \bar{F}_{a/m}^{(a)} \\ \bar{F}_{a/b}^{(a)} \\ M_{amx} \\ M_{amz} \\ M_{abx} \\ M_{aby} \end{bmatrix}_{12x1} = -\hat{F}_{12x2} \begin{bmatrix} \ddot{\psi}_{ac} \\ \ddot{\theta}_{mc} \end{bmatrix}_{2x1} + \hat{D}_{12x1} \qquad (13)$$

Using (13), the commanded torques, $T_{ac}$, $T_{ec}$, can be calculated when the reference values of angular accelerations, $\ddot{\psi}_{ac}$, $\ddot{\theta}_{mc}$, angular velocities, $\dot{\psi}_{ac}$, $\dot{\theta}_{mc}$, and angular positions, $\psi_{ac}$, $\theta_{mc}$, of the yaw and pitch gimbals are available. The reaction forces and moments, $X_i$, can again be calculated as by-products of the inverse model.

Complementary nature of the *Inverse* and the *Forward* models are shown in Fig. 6, where the outputs of the *Inverse Dynamics* model are the inputs of the *Forward Dynamics* model. Hence, by using the *Inverse Dynamics* model, it is possible to calculate the values of the required torque commands needed to achieve the prescribed reference values of angular accelerations, velocities and positions of the yaw and pitch gimbals. Moreover, the reaction forces and moments, $X_{iI}$, calculated from the inverse model and $X_{iF}$, calculated from the forward model are equal.

Note that, the position, velocity and acceleration outputs of the *Forward Dynamics* model are the same as the input position, velocity and acceleration commanded for the *Inverse Dynamics* model (double integral plant). If there exist different disturbances in *Inverse* and *Forward Dynamics*, (in terms of base platform motion, friction torque, model parameter uncertainties), double integral structure no longer exists. This feature helps to calculate the disturbance torque present in the real plant (Section III-IV).

The proposed gimbal model including *Inverse* and *Forward Dynamics* (with Eqns. 1-13) is implemented on the simulation environment by using MATLAB® [21] and Simulink® [27] software. Equations are programmed as MATLAB® Function blocks. In Fig. 7, simulation results are given for the z-component of the reaction moment on the revolute joint in between the inner and the outer gimbals of the two-axis gimbal, i.e., $M_{amz}$, generated from *Inverse* and *Forward Dynamics* models for the 2°@4Hz reference position input.

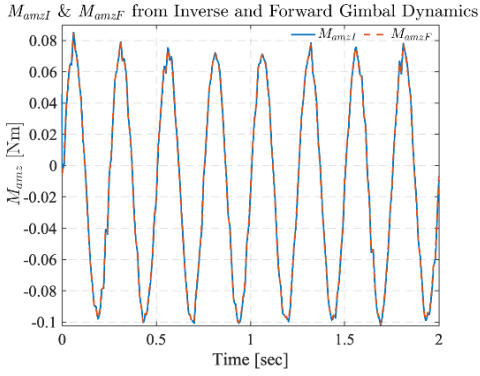

Fig. 7. z-component of $\bar{M}_{a/m}^{(a)}$ calculated from *Inverse* and *Forward Dynamics* model

## III. EXPERIMENTAL DATA COLLECTION

### A. THE TEST SET-UP

The experimental set-up used in this study is shown in Fig. 8. It includes a 2-DoF gimbal that is mounted on a Stewart platform with the red fixture, a data acquisition (DAQ) system (xPC Target and Simulink®), a host PC, gimbal electronic card box (including power, gimbal motor driving and image processing cards) and a power supply. Host PC and target PC communicate via the Ethernet communication protocol.

The values of the distances are given in Tab. B.I, Appendix B. The parameters associated with the gimbal platform are given in Tab. B.II and B.III for azimuth and elevation gimbals, in Appendix B. While calculating inertia and mass values, inner and outer gimbals are taken as separate rigid bodies. Motor parameters are taken from the datasheets of inner and outer gimbal motors.

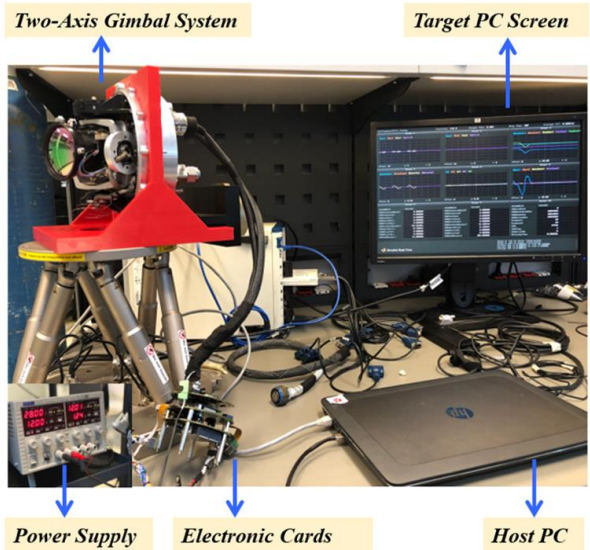

Fig. 8. The experimental set-up of the two-axis gimbal system

### B. DISTURBANCE TORQUE ESTIMATION

*Ideal Inverse Gimbal Dynamics* implies that there is no friction torque present in the revolute joint between the outer ring and the base platform ($T_{fra}$) and in the revolute joint between the inner and outer rings of the 2-DoF gimbal assembly ($T_{frm}$). There are no off-diagonal inertia parameters (or off-diagonal terms are very small compared to diagonal terms), rotation axis and CoG offsets, as well. In Fig. 9, the *Motor Dynamics* block between *Ideal Inverse Gimbal Dynamics* and *Real Plant* converts Torque to Voltage (T2V) using the data from Appendix B, Tab. B.II. Commanded acceleration ($\ddot{\bar{r}}_d$), velocity ($\dot{\bar{r}}_d$) and position ($\bar{r}_d$) are the inputs for *Ideal Inverse Gimbal Dynamics*. The output of the *Inverse Gimbal Dynamics* block, is the required torque ($\bar{u}_d = [T_{ac} \quad T_{mc}]^T$) needed to perform the commanded motion for an ideal system. Real plant outputs are encoder and gyro data ($\bar{r}$, $\dot{\bar{r}}$). To derive the acceleration input, $\ddot{\bar{r}}$, a backwards derivative operation at every 10 data points is performed for the gyro data in xPC Target. Real plant input is the voltage, $\bar{V}_c$. It has been observed that, open loop response of the system changes under different sinusoidal inputs (i.e., system produces different oscillation patterns in position and velocity). Due to disturbance torques (especially cable restraint torques and coupling), *the actual plant* response cannot catch the magnitude of the sinusoidal reference, does not oscillate around 0° and attain a different shape from a sine wave. In order to decide on the control strategy that will be applied for the real plant, it is important to understand the nature of the disturbance torque.

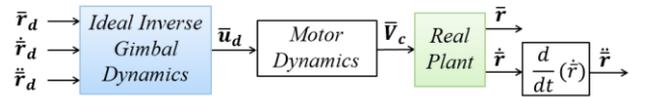

Fig. 9. Open loop block diagram used in xPC Target for obtaining *Real Plant* outputs

Multi-axis gimbal is a chain mechanism. Linkages are connected by joints to build up the mechanism. Thus, there exists disturbance torques resulting from assembling the linkages (see Section II, Eqns. 12, 13) and they are dependent on the model parameters (distances and inertia terms). For this reason, overall disturbance acting on the system cannot be represented by a general friction model. Based on our detailed mathematical model presented in Section II, disturbance torque can be estimated by using the *block diagram structure* in Fig. 10. It is assumed that the differential torque between the first and second *Ideal Inverse Dynamics* blocks is the disturbance torque. Similar to Fig. 9, in order to derive the acceleration input, $\ddot{\bar{r}}$, that is used as an input for the second *Ideal Inverse Gimbal Dynamics* in the block diagram of the real system, a backwards derivative operation at every 10 data points is performed for the gyro data. The output of the second *Inverse Gimbal Dynamics* block, is the corresponding torque ($\bar{u}'_d = [T_{ma} \quad T_{mm}]^T$) produced from the output of the *Real Plant* (set-up in Fig. 8). $\overline{\Delta u} = [T_{da} \quad T_{dm}]^T$ is the disturbance torque (differential torque) in the system.

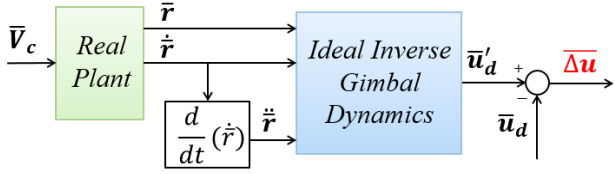

Fig. 10. Open loop block diagram used in xPC Target for calculating $T_{da}$ and $T_{dm}$

The disturbance torque present in the system depends on states $(\theta_m, \dot{\theta}_m, \psi_a, \dot{\psi}_a)$. There is coupling between azimuth and elevation gimbals, so that the motion of one gimbal affects the other. In (14), $(x = a, m)$, the terms in the disturbance function are assume to be single and dual combinations of $\theta_m, \dot{\theta}_m, \psi_a, \dot{\psi}_a, \theta_m^2, \psi_a^2, \dot{\theta}_m^2, \dot{\psi}_a^2$ and there is the bias term $(k37_x)$.

The parameters $k1_x, ..., k36_x$ are associated with the state dependent friction torques, cogging torques of wires and cooling pipes. Similarly, the parameter $k37_x$ is associated with the state independent bias terms, disturbances related to mechanism and linkages (i.e., off-diagonal inertia terms and CoG, rotation axis offsets).

$$T_{dx} = k1_x\dot{\theta}_m + k2_x\theta_m + \cdots + k34_x(\dot{\psi}_a^2\theta_m^2) + k35_x(\dot{\psi}_a^2\dot{\theta}_m^2) + k36_x(\dot{\theta}_m^2\theta_m^2) + k37_x \quad (14)$$

In order to identify the coefficients in $T_{da}$ and $T_{dm}$, several randomly generated datasets are collected from the experimental set-up: (1) $10°@2Hz$ position input is commanded for the azimuth gimbal and $7°@2Hz$ position input is commanded for the elevation gimbal, (2) $3°@5Hz$ position input is commanded both for the azimuth gimbal and the elevation gimbal, (3) $5°@4Hz$ position input is commanded both for the azimuth gimbal and the elevation gimbal. $T_{ac}, T_{ma}$ and $T_{da}$ are plotted for the azimuth gimbal in Fig. 11 for (1).

Collected data is used to form an equation of the form, $A\bar{x} = \bar{b}$, which is solved in the least sqaure sense. Corresponding to each dataset, a set of coefficients are identified. Symbolic representations of matrix **A** and vectors $\bar{x}$ and $\bar{b}$ are given in (15), (16) and (17).

$$A = \begin{bmatrix} \dot{\theta}_m & \theta_m & \cdots & 1 & 0 & 0 & \cdots & 0 \\ 0 & 0 & \cdots & 0 & \dot{\theta}_m & \theta_m & \cdots & 1 \end{bmatrix} \quad (15)$$

$$\bar{x} = [k1_a \quad k2_a \quad \cdots \quad k37_a \quad k1_e \quad k2_e \quad \cdots \quad k37_e]^T \quad (16)$$

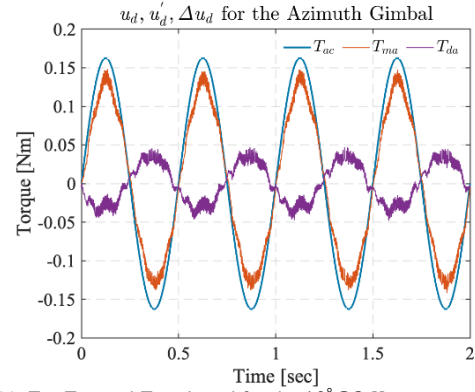

Fig. 11. $T_{ac}, T_{ma}$ and $T_{da}$ plotted for the $10°@2 Hz$ commanded position input of the azimuth gimbal

$$\bar{b} = \begin{bmatrix} T_{da} \\ T_{dm} \end{bmatrix} \quad (17)$$

In a linear system, it is expected to find similar coefficients from each dataset and form a single function that represents the disturbance torque for the whole operational range. However, each coefficient set is very different from each other. For instance, Table I shows the coefficients $k1_a$ and $k37_a$ calculated from each dataset. Each coefficient set works well for the dataset that is calculated from; but fails to represent the system when it is applied to the other dataset. Fig. 12(a) shows the azimuth gimbal's estimated disturbance torque calculated with the coefficients found from (1) for the disturbance torque present in (1). Fig. 12(b) shows the azimuth gimbal's estimated disturbance torque calculated with the coefficients found from (2) for the disturbance torque present in (1).

TABLE I
Coefficients from Datasets (1), (2) and (3)

| ∴ | (1) | (2) | (3) |
|---|---|---|---|
| $k1_a$ ($\frac{Nms}{rad}$) | 0.036 | 2.201 | 0.281 |
| $k37_a$ ($Nm$) | -0.094 | -10.944 | 3.895 |

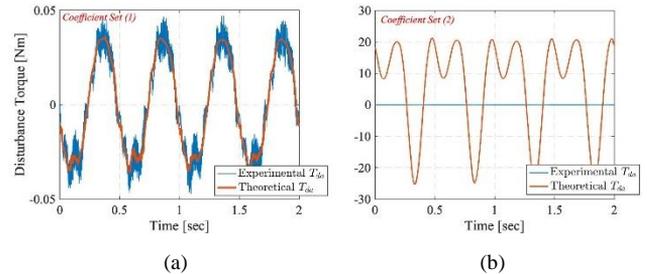

Fig. 12. The estimated disturbance torque-real disturbance torque graph plotted for the azimuth gimbal with coefficient set (1) and (2)

These observations lead us to the conclusion that there exists a state dependent, highly nonlinear and complex disturbance torque present in the experimental set-up. So, it is very difficult for us to fit a general disturbance model that represents the behavior in the FOR range of the setup. Instead, we propose a **nonlinear regressor, "a multi-layer perceptron"**, to represent the disturbance torque in the system and perform torque compensation.

Disturbance torque functions for azimuth and elevation axes used in simulations in Section IV have a complex and highly nonlinear form given in (18) and (19), (the terms with relatively small coefficients are eliminated from the general disturbance function proposed in (14)).

$$T_{da} = -0.08\dot{\theta}_m - 0.01\theta_m - 0.02\theta_m^2 - 0.2\dot{\psi}_a - 0.01\psi_a \\ - 0.01\theta_m\psi_a - 0.02\theta_m^3 - 0.1\theta_m\psi_a^2 \\ - 0.05\psi_a\theta_m^2 - 0.00045 \quad (18)$$

$$T_{dm} = -0.08\dot{\theta}_m - 0.05\theta_m - 0.08\theta_m^2 - 0.02\dot{\psi}_a - 0.01\psi_a \\ - 0.02\psi_a^2 - 0.03\theta_m\psi_a - 0.02\theta_m\psi_a^2 \\ - 0.02\psi_a\theta_m^2 - 0.00045 \quad (19)$$

## IV. MLP-ASSISTED CONTROL STRUCTURE

In this section, neural network-based control structure is developed for a two-axis gimbal system. Simulations of the system with the NN-based controller, cascaded PID controller and ADRC with linear ESO are carried out in Simulink®. Then, the proposed controller is tested in the real set-up shown in Section III by using Simulink Real-Time®.

### A. NN BASED CONTROL STRUCTURE WITH INVERSE AND FORWARD GIMBAL DYNAMICS (TORQUE COMPENSATION)

This section is about disturbance torque compensation by using *"Ideal Inverse Dynamics-Disturbed Forward Dynamics-Ideal Inverse Dynamics"* sequence which is implemented in MATLAB®. *The actual gimbal plant* is replaced by *Disturbed Forward Gimbal Dynamics* in simulations. NN training is performed with the block diagram given in Fig. 13.

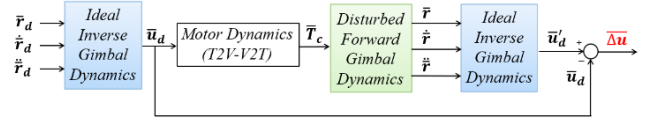

Fig. 13. Training structure for NN used in Simulink®

When desired acceleration ($\ddot{\bar{r}}_d$), velocity ($\dot{\bar{r}}_d$) and position ($\bar{r}_d$) are feed into *Ideal Inverse Gimbal Dynamics*, required torque ($\bar{u}_d$) needed to perform this motion will be produced for an ideal system. When $\bar{u}_d$ is feed into *Disturbed Forward Gimbal Dynamics or Real Plant*, resultant motion will be equal ($r, \dot{r}, \ddot{r}$) which is different than ($\bar{r}_d, \dot{\bar{r}}_d, \ddot{\bar{r}}_d$) due to disturbance torque present in the system. $\bar{u}'_d$ is the torque value corresponding to motion of the *Real Plant* or the representative plant of the real system in the simulations ($r, \dot{r}, \ddot{r}$). It is obtained by sending the output of *Disturbed Forward Gimbal Dynamics (Real Plant)* to *Ideal Inverse Gimbal Dynamics*. The training data set of the neural network is $\{(\bar{r}_d, \dot{\bar{r}}_d, \bar{u}_d), \overline{\Delta u}, t \in [0, t_f]\}$. Output of the NN is the disturbance torque ($\overline{\Delta u}$).

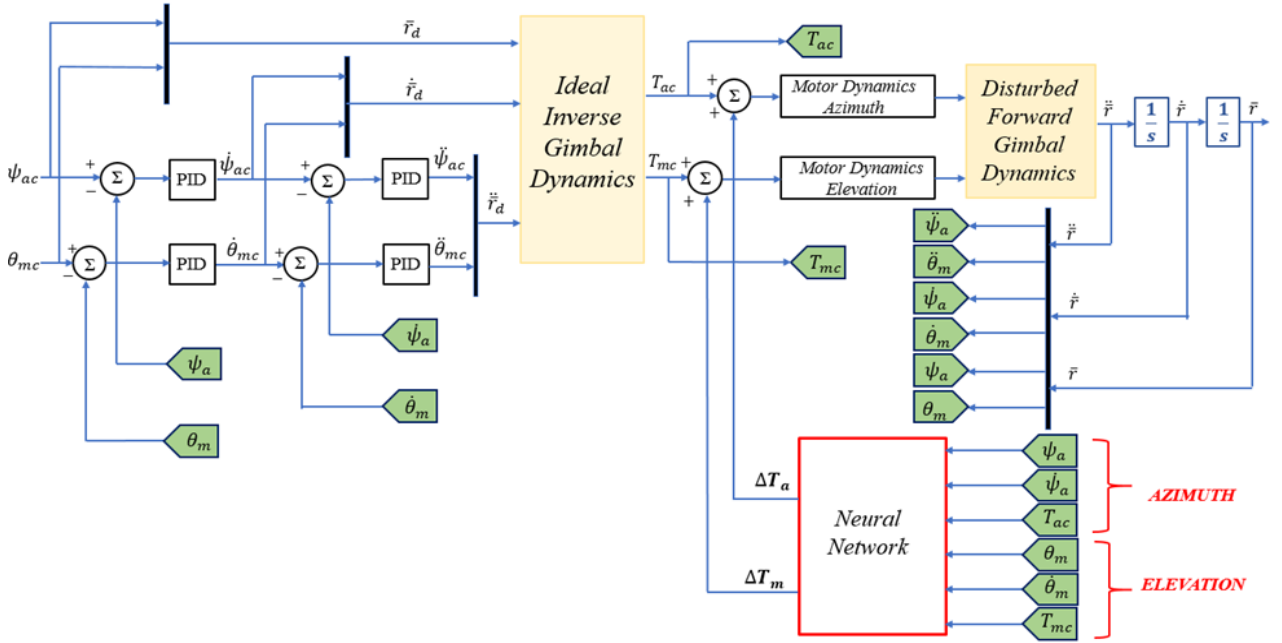

Fig. 14. Block diagram implementation of NN with cascaded PID controllers in Simulink®

The proposed design procedure of the NN-based control strategy is presented in the block diagram in Fig. 14 for the simulations. Note that, acceleration ($\ddot{\bar{r}}$) is the output of the *Disturbed Forward Dynamics* block. The difference between $\bar{u}'_d$ and $\bar{u}_d$ results in *Delta Torque* ($\overline{\Delta u}$), enables the real plant to reach the reference position in a closed loop structure when it is added to $\bar{u}_d$ (Fig. 14). It works as a disturbance torque compensator that compensates the loss in the plant. It also works as a feedforward torque estimator; because torque output of PID controller *(Ideal Inverse Gimbal Dynamics)* enters the NN. It enhances the response of the system when the PID controller is not sufficient enough to perform the desired motion (improves target tracking performance).

In the examples given below, NN is trained with an amplitude decreasing, frequency increasing sinusoidal wave. Training input changes from 30° to 1°, for the azimuth axis and it changes from 15° to 1°, for the elevation axis with frequency starting from 0.5Hz and ending at 5Hz. With this input, almost all ranges in the FOR limits of the experimental set-up are covered.

The feedforward neural network that is designed for the simulations has 2 hidden layers, each composed of 20 neurons. The training method is chosen as Levenberg-Marquardt backpropagation. In one scenario, after gimbal reaches the target and get locked onto it, it is expected to perform high frequency and low amplitude motion. Thus, PID controllers are ideally tuned under a pulse input with ±3° amplitude, 2 second period and 50% pulse width. The response of the cascaded PID controller for the tuning input is given in Fig. 15 for the azimuth gimbal.

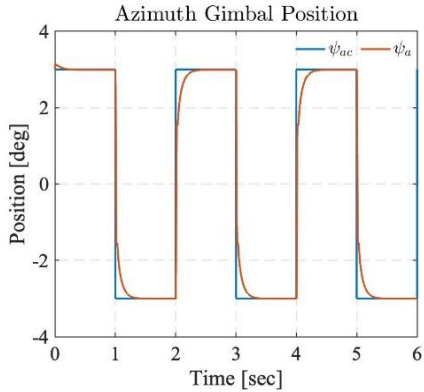

Fig. 15. Response of the azimuth gimbal for the pulse input with PID controller

During the search pattern (while gimbal is trying to locate the position of the target), it may be required for the gimbal to oscillate with larger amplitudes and slower frequencies. Response of the system with NN controller is plotted under 5°@1Hz sinusoidal reference for the azimuth gimbal in Fig. 16. Without NN, it is observed that azimuth gimbal reaches to 4.75°. ***NN helps the gimbal to reach the desired position as well as it speeds up the response, decreases time delay.***

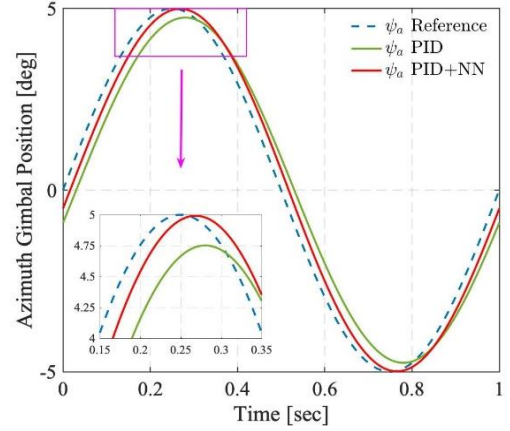

Fig. 16. Azimuth gimbal position for the sinusoidal input with PID and NN controllers

NN based control structure increases the bandwidth and gain margin of the system. By applying the chirp signal that reaches 20Hz in 10 seconds with amplitude 1° for azimuth gimbal, the bandwidth is found. Azimuth gimbal reaches to 70.7% of the reference input around 5Hz (2.5th second) as shown in Fig. 17(a). First 5 seconds of the chirp signal and response are shown in the plot. It is observed that the response of the system that only runs with PID controllers is around 0.73°, whereas response of the system that runs with PID + NN is better. Azimuth gimbal reaches to 0.88° in Fig. 17(b). Motor current applied to the system is within the saturation limits in both cases. NN based controller, increases bandwidth of the azimuth gimbal approximately 4.2 times for the sinusoidal wave with 1° amplitude.

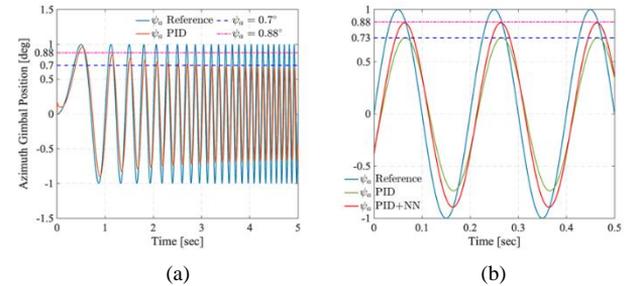

(a) (b)

Fig. 17. Azimuth gimbal response corresponding to chirp signal (a), 1°@ 5Hz input (b)

## B. ADRC IMPLEMENTATION FOR INVERSE AND FORWARD GIMBAL DYNAMICS

Active Disturbance Rejection Control (ADRC) theory is invented by J. Han in [29]. It is a method to replace the conventional PID controller. The block diagram implementation of ADRC with *Inverse and Forward Dynamics* is shown in Fig. 18.

Instead of using a NN based controller, disturbance torque compensation can also be supplied from using *Ideal Inverse Dynamics*.

10                                                                                                                                     DRAFT

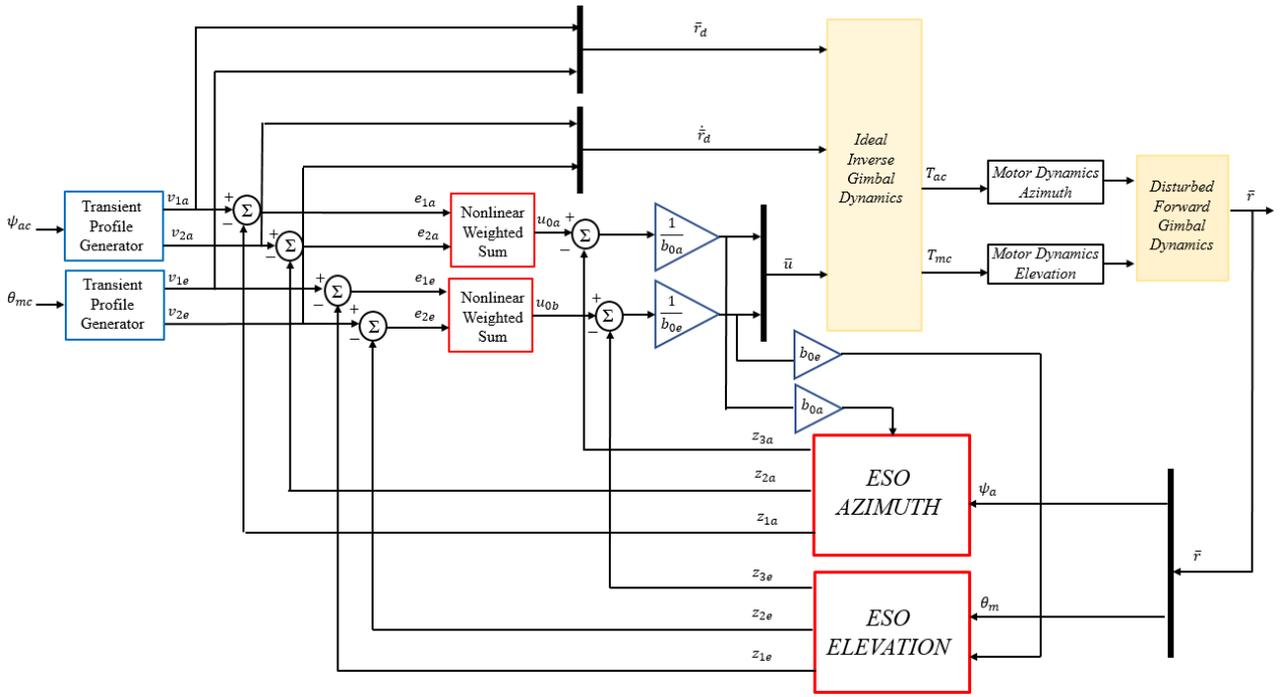

Fig. 18. Block diagram implementation of ADRC with *Inverse* and *Forward Gimbal Dynamics* in Simulink®

By sending the output of *Disturbed Forward Gimbal Dynamics* to the second *Ideal Inverse Gimbal Dynamics* block, the differential torque between the first *Ideal Inverse Gimbal Dynamics* and the *Second Ideal Inverse Gimbal Dynamics* can be calculated and added to system. In other words, with this approach, instead of training a NN with the data obtained from *Ideal Inverse Dynamics,* the training data can be directly supplied from *Ideal Inverse Dynamics.* However, this approach is not applicable as wide as the NN-controller suggested in this study, especially for relatively high frequency reference inputs (Figs. 21, 22). The block diagram implementation of cascaded PID controllers with *Inverse Dynamics* based compensation is shown in Fig. 19.

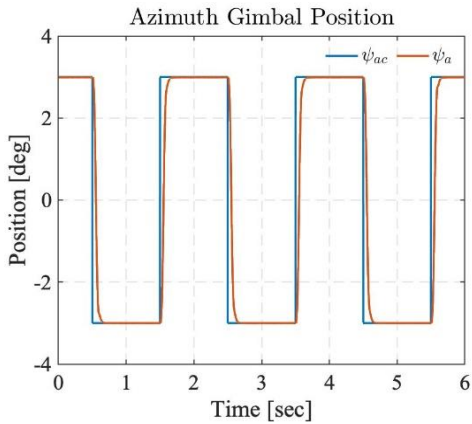

Fig. 20. Response of the azimuth gimbal for the pulse input with ADRC controller

In the following examples, comparisons between PID, NN-based controller, ADRC with ESO and inverse model based controller are provided for different reference inputs. Amplitude of the reference is held constant at 5°. Frequency of the sinusiodal wave is varied from 1Hz to 5 Hz (Reference Sets 1-5). The ADRC controller is optimally tuned under a pulse input with ±3° amplitude, 2 seconds period and 50% pulse width. The response of the ADRC controller for this input is given in Fig. 20 for the azimuth gimbal.

Figs. 21 and 22 show the response of the aforementioned controllers for Reference Sets 1 and 5 for the azimuth gimbal. Tabs. II and III give the percent improvement of mean tracking and peak errors wrt. PID controller. Percent decrease in Tabs. II and III indicates the percent how much each controller decreases mean tracking and peak errors with respect to (wrt.) PID controller. Fig. 23 shows the change in these errors for each controller corresponding to references 1Hz-5Hz.

TABLE II
Percent Improvement (Azimuth) wrt. PID Controller for Reference Set 1

|  | *Percent Decrease (%)* | |
| --- | --- | --- |
|  | *Mean Track Error* | *Peak Error* |
| **PID+NN** | 44.3 | 97.65 |
| **ADRC** | 44.2 | 88.13 |
| **Inverse Model** | 31.54 | 60.57 |

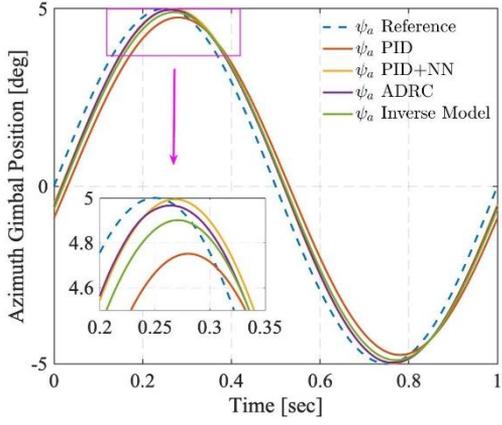

Fig. 21. Response of the azimuth gimbal for Reference Set 1

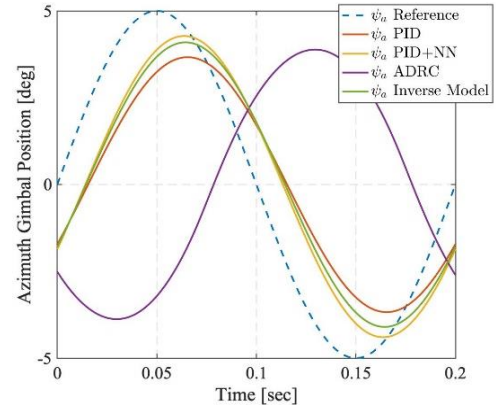

Fig. 22. Response of the azimuth gimbal for Reference Set 5

TABLE III
Percent Improvement (Azimuth) wrt. PID Controller for Reference Set 5

|  | Percent Decrease (%) | |
| --- | --- | --- |
|  | *Mean Track Error* | *Peak Error* |
| **PID+NN** | 14.25 | 54.5 |
| **ADRC** | -246.48 | 15.8 |
| **Inverse Model** | 31.54 | 60.57 |

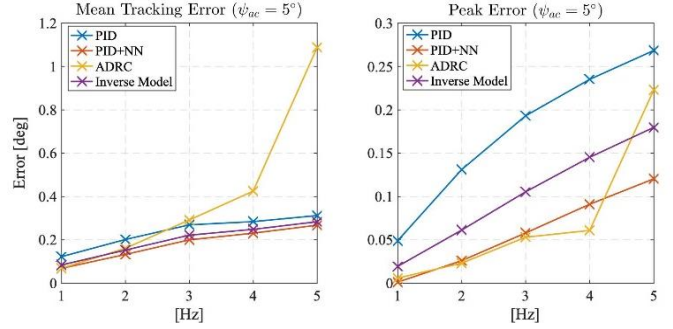

Fig. 23. Mean tracking and peak errors (azimuth) of each controller for Reference Sets 1-5

Among cascaded PID, NN-based, ADRC and inverse model based controllers, according to mean and peak value figures and tables, best controllers are ADRC ve NN-based controllers. In terms of peak error, ADRC controller is slightly better than the PID controller in the low frequency range for the azimuth axis (2-3Hz references); but there is no pattern in it. ADRC is input dependent. Based on the reference set, ADRC can perform better/worse than the simple PID controller in terms of tracking error. NN-based controller is always better than the cascaded PID.

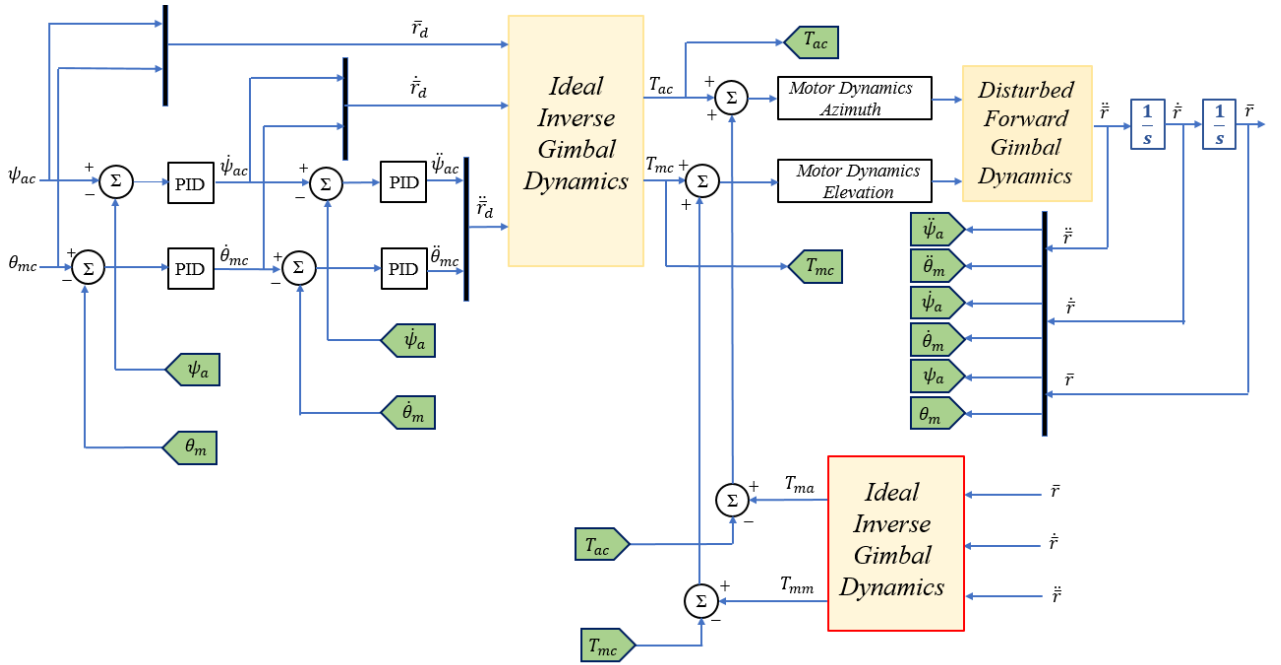

Fig. 20. Block diagram implementation of cascaded PID controllers with *Inverse Dynamics* based compensation in Simulink®

***ADRC requires tuning and design effort for different reference sets (Fig. 23).*** It is difficult and it takes time to tune the coefficients of ADRC controller. Disturbance torques present in the system are coupled coming from two different axes. This is verified by the data collected from the experimental set-up in Section III. **ADRC controller uses distinct ESOs for each channel.** Instead, in the proposed algorithm, there is only one NN and it works as a MIMO disturbance torque compensator. Inputs of the NN come from both the azimuth and elevation axes.

PID and ADRC controllers are designed optimaly. On the other hand, one cannot say NN being optimal. It depends on the training dataset, layer and neuron numbers, training algorithm. So, it has a chance to improve its performance further. Especially, training set has an important effect on the performance of the NN in our system. ADRC is basically applicable for double integral systems. In our system, *Inverse Dynamics* is ideal; however there is disturbance in the *Forward Plant*. So, the double integral structure is in a way lost. This might be another reason why ADRC controller does not work as well as NN based controller in our system.

### C. IMPLEMENTATION OF NN BASED CONTROLLER FOR REAL-TIME EXPERIMENTS

Proposed control strategy is tested in the experimental set-up by using xPC Target. Block diagram representation given in Fig. 14 is implemented on the host PC. Input for the real system is $\bar{V}_c$. During experiments, gyro data is used by passing it through a 100 Hz low-pass filter. While obtaining the training data for the NN, acceleration is derived by using the filtered version of the gyro data and taking derivative at every 100 points. NN structure used in the experiments is same as the ones used in Simulink®.

In Example 1, NN is trained with the inputs given in Fig. 24 (red crosses), sine waves are used for the azimuth gimbal and cosine waves are used for the elevation gimbal.

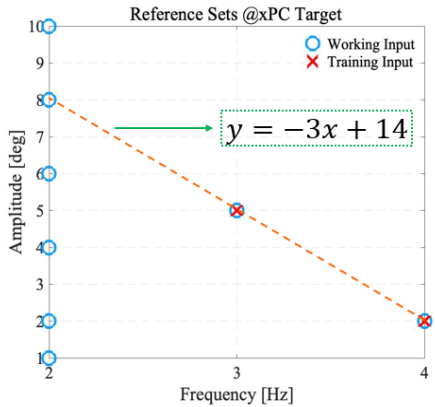

Fig. 24. Reference sets used with the physical gimbal for Example 1

It is observed that the response of the system that runs with PID + NN is better compared to response of the system that runs only with PID controllers for the training input (Fig. 24, red crosses) and for the input that is outside the range of training set (Fig. 24, blue circles). The orange line in Fig. 24, corresponds to $y = -3x + 14$ equation. This is the reason 8°@2Hz and other amplitudes around this reference input are choosen to be studied with the NN traned in Example 1.

Fig. 25, shows the response of the azimuth gimbal for the training input (5°@3Hz). Fig. 26 shows the response of the azimuth and elevation gimbals for the reference input (6°@2Hz).

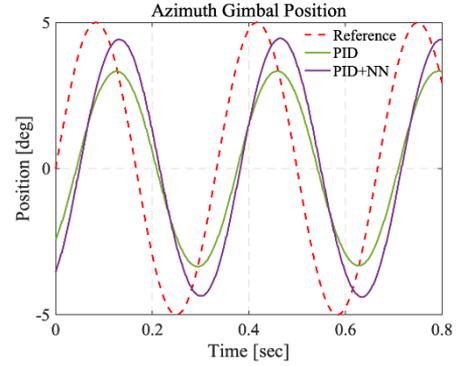

Fig. 25. Response of the azimuth gimbal for the training input (5°@3Hz) with NN based controller in xPC Target

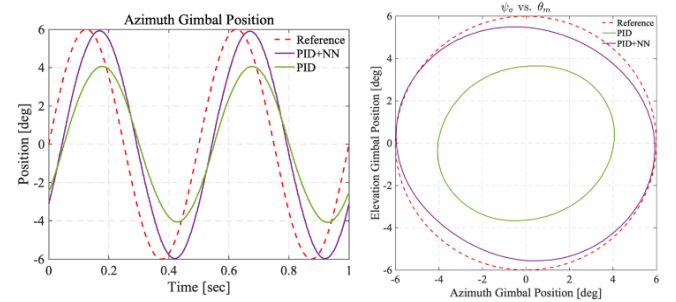

Fig. 26. Response of the azimuth gimbal for the 6°@2Hz reference with NN based controller in xPC Target

In Example 2, NN is trained and tested with the inputs 2,3°@5Hz sine waves for the azimuth and 2,3°@5Hz cosine waves for the elevation gimbals. By applying chirp signal for azimuth and elevation gimbals with amplitude 2°, the bandwidth of the set-up is found (Fig. 27). Elevation gimbal response corresponding to chirp signal is given in Fig. 27 (chirp signal reaches 20Hz in 10 seconds with amplitude 2°). Elevation gimbal reaches to 1.9° approximately around 3 Hz and to 1.5° around 5Hz. First 6 seconds of the chirp signal and response are shown in the plot.

In Fig. 27, as the frequency increases response of the system tends to be more anti-symmetric wrt. x-axis and shows a different oscillation pattern in terms of shape and magnitude. This is the implication of having a highly nonlinear system. Response of the elevation gimbal for the 2°@5Hz cosine wave is shown in Fig. 28. NN based controller can reach to 1.95° and PID controller can reach to 1.55° around 5Hz. NN based controller increases the bandwidth of the system approximately around 66%. This NN, besides increasing the amplitude of the response, also decreases the phase difference for this reference.

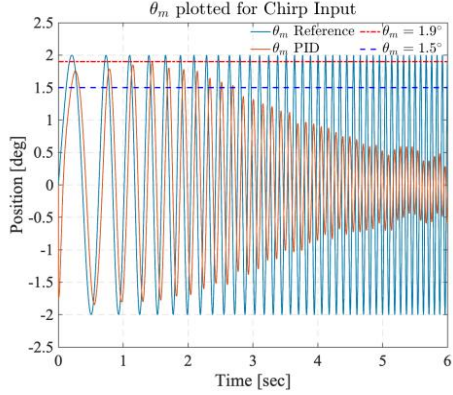

Fig. 27. Elevation gimbal response corresponding to chirp signal in xPC Target with PID controllers

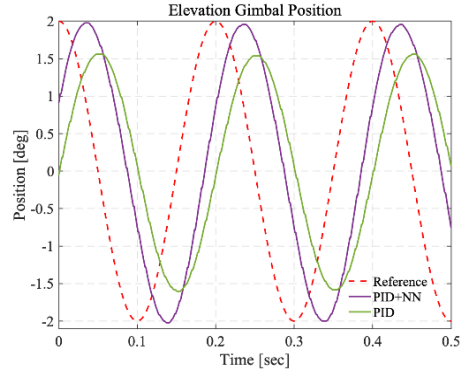

Fig. 28. Elevation gimbal response corresponding to 2°@5Hz reference with NN based controller in xPC Target

### D. EFFECT OF THE INVERSE MODEL

In this study, a novel training method for the NN based on the sequential use of *Inverse* and *Forward Dynamics* is proposed. While working with the experimental set-up, it is essential to train the NN with the most realistic *Inverse Dynamics Model* as possible. By using a detailed *Inverse Model,* it is more likely to capture the effects of total disturbances on the dynamical behavior of the system (including the effects of CoG offsets, rotation axis misalignments, dynamical mass unbalance and friction). In Fig. 29, comparison is provided for the performances of NNs trained with the *Inverse Dynamics* proposed in Section II and with the *Simple Inverse Dynamics* that is derived using the Eqns. of the *Forward Dynamics Models* in [12], [14], [15], (gimbal dynamics is basically represented with a rotating inertia term).

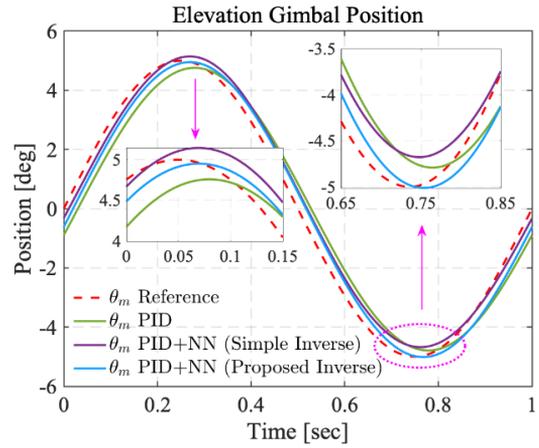

Fig. 29. Elevation gimbal response for NNs trained with *Simple* and *Proposed Inverse Dynamics Models*

## V. CONCLUSIONS AND FUTURE WORK

In this article, a novel and fully detailed dynamical model for a two-axis gimbal system is proposed. Different from the articles, EOMs of the system are derived using multi-body dynamics approach. Mathematical model proposed in this study covers the misalignments between rotation axes of gimbals, the offsets between the CoG of the gimbals and the rotation axes, dynamical mass unbalance and the disturbances on the joints arising from friction or restraining elements of the whole gimbal system. Both translational and rotational motions are considered. Furthermore, by using Newton-Euler approach, reaction forces and moments on the joints of the gimbal system are expressed. After the derivation of the *Forward Dynamics* model, the *Inverse Dynamics* model is generated to estimate the complex, nonlinear, state and mechanism dependent disturbance torques present in the system. By using *Inverse* and *Forward Gimbal Dynamics* consecutively, a new procedure for collecting the training data of a NN is developed. After the training, NN-assisted control structure is able to express the real physical phenomena affecting the real system exactly and improves the response almost in the full operational range (decreases mean tracking error up to 44.6% compared to PID).

As a future research, online training of the NN will be implemented on the physical set-up. A parameter-based performance analysis of the NN-based controller will be conducted.

14                                                                                                                                                                     DRAFT

## VI. APPENDIX A

Coordinate frame transformations and *Kinematic Equations for the Rotational Motion* of the two-axis gimbal are presented in this section. In Fig. A.1, the complete rotational motion of the two-axis system is shown in terms of three rotation motion. The rotation between the inertial reference frame $F_o$ (with axes $x_o, y_o, z_o$) and the base platform reference frame $F_b$ (with axes $x_b, y_b, z_b$) is defined using a widely used 3-2-1 (yaw-pitch-roll) rotation sequence.

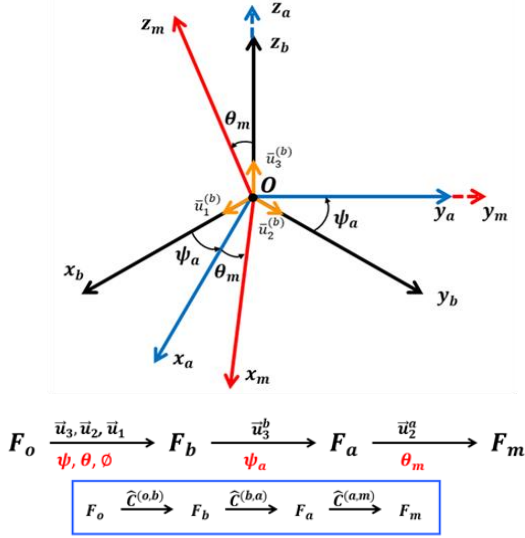

Fig. A.1. Euler angles between reference frames ($\bar{u}_1^{(i)}, \bar{u}_2^{(i)}, \bar{u}_3^{(i)}$ are the unit vectors along $x_i, y_i, z_i$ directions for reference frames $i = b, a, m$)

The angular velocity ($\bar{\omega}_{b/o}^{(b)}$) and angular acceleration ($\bar{\alpha}_{b/o}^{(b)}$) of the base platform with respect to a chosen inertial reference frame, expressed in the base reference frame and denoted in column matrix representation are given as:

$$\bar{\omega}_{b/o}^{(b)} = [p\ q\ r]^T, \quad \bar{\alpha}_{b/o}^{(b)} = [\dot{p}\ \dot{q}\ \dot{r}]^T \tag{A1}$$

The angular velocity ($\bar{\omega}_{a/o}^{(a)}$) and angular acceleration of the outer gimbal ($\bar{\alpha}_{a/o}^{(a)}$) with respect to a chosen inertial reference frame expressed in the outer gimbal reference frame is also presented below in (A2).

$$\bar{\omega}_{a/o}^{(a)} = \dot{\psi}_a \bar{u}_3^{(a)} + \hat{R}_3(-\psi_a)\bar{\omega}_{b/o}^{(b)} \tag{A2}$$

$$\bar{\alpha}_{a/o}^{(a)} = \ddot{\psi}_a \bar{u}_3^{(a)} + D\bar{\alpha}_{a/o}^{(a)} \tag{A3}$$

$$D\bar{\alpha}_{a/o}^{(a)} = \hat{R}_3(-\psi_a)\bar{\alpha}_{b/o}^{(b)} - \dot{\psi}_a \tilde{u}_3 \hat{R}_3(-\psi_a)\bar{\omega}_{b/o}^{(b)} \tag{A4}$$

Here, $\bar{u}_i^{(.)}(i = 1,2,3)$ is the unit vector of the reference frame denoted with $(.)$. Also, the Tilde operator $(\sim)$

LEBLEBICIOGLU ET AL.

constructs the skew symmetric matrix form of the column matrix representation of a vector. $D\bar{\alpha}_{a/o}^{(a)}$ in (A4) is used to shorten the expression for $\bar{\alpha}_{a/o}^{(a)}$ in (A3).

The angular velocity ($\bar{\omega}_{m/o}^{(m)}$) and angular acceleration of the pitch gimbal ($\bar{\alpha}_{m/o}^{(m)}$) with respect to a chosen inertial reference frame expressed in the inner gimbal reference frame is presented below. Here, $D_1\bar{\alpha}_{m/o}^{(m)}$ and $D_2\bar{\alpha}_{m/o}^{(m)}$, (Eqns. A7 and A8), are used to shorten the expression for $\bar{\alpha}_{m/o}^{(m)}$ in (A6).

$$\bar{\omega}_{m/o}^{(m)} = \dot{\theta}_m \bar{u}_2^{(m)} + \dot{\psi}_a \hat{R}_2(-\theta_m)\bar{u}_3^{(a)} + \hat{R}_2(-\theta_m)\hat{R}_3(-\psi_a)\bar{\omega}_{b/o}^{(b)} \tag{A5}$$

$$\bar{\alpha}_{m/o}^{(m)} = \ddot{\theta}_m \bar{u}_2^{(m)} + \ddot{\psi}_a D_1 \bar{\alpha}_{m/o}^{(m)} + D_2 \bar{\alpha}_{m/o}^{(m)} \tag{A6}$$

where,
$$D_1\bar{\alpha}_{m/o}^{(m)} = \hat{R}_2(-\theta_m)\bar{u}_3^{(a)} \tag{A7}$$

and,
$$\begin{aligned}D_2\bar{\alpha}_{m/o}^{(m)} =\ & \hat{R}_2(-\theta_m)\hat{R}_3(-\psi_a)\bar{\alpha}_{b/o}^{(b)} \\ & - \dot{\psi}_a \hat{R}_2(-\theta_m)\hat{R}_3(-\psi_a)\bar{\omega}_{b/o}^{(b)} \\ & - \dot{\theta}_m \dot{\psi}_a \tilde{u}_2 \hat{R}_2(-\theta_m)\bar{u}_3^{(a)} \\ & - \dot{\theta}_m \tilde{u}_2 \hat{R}_2(-\theta_m)\hat{R}_3(-\psi_a)\bar{\omega}_{b/o}^{(b)}\end{aligned} \tag{A8}$$

The expressions used in order to shorten the Eqns. (3) and (6), are given in explicit form below:

$$\bar{D}a_{Ga}^{(a)} = [\widetilde{D\bar{\alpha}_{a/o}^{(a)}} + (\widetilde{\omega}_{a/o}^{(a)})^2]\bar{r}_{Ga/a}^{(a)} + \hat{R}_3(-\psi_a)[\tilde{\alpha}_{b/o}^{(b)} + (\widetilde{\omega}_{b/o}^{(b)})^2]\bar{r}_{a/b}^{(b)} + \hat{R}_3(-\psi_a)\bar{a}_{b/o}^{(b)} \tag{A9}$$

$$\begin{aligned}\bar{D}a_{Gm}^{(m)} =\ & [\widetilde{D_2\bar{\alpha}_{m/o}^{(m)}} + (\widetilde{w}_{m/o}^{(m)})^2]\bar{r}_{Gm/m}^{(m)} \\ & + \hat{R}_2(-\theta_m)\hat{R}_3(-\psi_a)[\tilde{\alpha}_{b/o}^{(b)} \\ & + (\widetilde{\omega}_{b/o}^{(b)})^2]\bar{r}_{a/b}^{(b)} \\ & + \hat{R}_2(-\theta_m)\hat{R}_3(-\psi_a)\bar{a}_{b/o}^{(b)} \\ & + \hat{R}_2(-\theta_m)[\widetilde{D\bar{\alpha}_{a/o}^{(a)}} + (\widetilde{w}_{a/o}^{(a)})^2]\bar{r}_{m/a}^{(a)}\end{aligned} \tag{A10}$$

*Matrices Used in Forward and Inverse Dynamic Models:*

The matrices, $\hat{F}, \hat{R}, \hat{D}$ and $\hat{G}$ (in Eqns. 12, 13) are given below in (A11)-(A14), respectively.

$$\hat{F} = \begin{bmatrix} m_a \tilde{u}_3 \bar{r}_{Ga/a}^{(a)} & \bar{0} \\ \hat{J}_a \bar{u}_3 & \bar{0} \\ m_m[\hat{R}_2(-\theta_m)\tilde{u}_3 \bar{r}_{m/a}^{(a)} + \widetilde{D_1\bar{\alpha}_{m/o}^{(m)}}\bar{r}_{Gm/m}^{(m)}] & m_m \tilde{u}_2 \bar{r}_{Gm/m}^{(m)} \\ \hat{J}_m D_1 \bar{\alpha}_{m/o}^{(m)} & \hat{J}_m \bar{u}_2 \end{bmatrix} \tag{A11}$$

$$\hat{R} = \begin{bmatrix} -\hat{I}_{3x3} & -\hat{I}_{3x3} & \hat{0}_{3x2} & \hat{0}_{3x2} \\ -\tilde{r}_{a/m}^{(a)} & -\tilde{r}_{a/b}^{(a)} & -\begin{bmatrix} 1 & 0 \\ 0 & 0 \\ 0 & 1 \end{bmatrix} & -\begin{bmatrix} 1 & 0 \\ 0 & 1 \\ 0 & 0 \end{bmatrix} \\ \hat{R}_2(-\theta_m) & \hat{0}_{3x3} & \hat{0}_{3x2} & \hat{0}_{3x2} \\ \tilde{r}_{m/a}^{(m)}\hat{R}_2(-\theta_m) & \hat{0}_{3x3} & \hat{R}_2(-\theta_m)\begin{bmatrix} 1 & 0 \\ 0 & 0 \\ 0 & 1 \end{bmatrix} & \hat{0}_{3x2} \end{bmatrix} \quad (A12)$$

$$\hat{D} = \begin{bmatrix} -m_a \bar{D} a_{Ga}^{(a)} + m_a \bar{g}_a \\ -\hat{J}_a D \bar{\alpha}_{a/o}^{(a)} - \tilde{\omega}_{a/o}^{(a)} \hat{J}_a \bar{\omega}_{a/o}^{(a)} + \begin{bmatrix} 0 \\ -T_{frm} \\ -T_{fra} \end{bmatrix} \\ -m_m \bar{D} a_{Gm}^{(m)} + m_m \bar{g}_m \\ -\hat{J}_m D_2 \bar{\alpha}_{m/o}^{(m)} - \tilde{\omega}_{m/o}^{(m)} \hat{J}_m \bar{\omega}_{m/o}^{(m)} - \hat{R}_2(-\theta_m)\begin{bmatrix} 0 \\ -T_{frm} \\ 0 \end{bmatrix} \end{bmatrix} \quad (A13)$$

$$\hat{G} = \begin{bmatrix} \hat{0}_{3x2} \\ \begin{bmatrix} 0 & 0 \\ 0 & 1 \\ 1 & 0 \end{bmatrix} \\ \hat{0}_{3x2} \\ -\hat{R}_2(-\theta_m)\begin{bmatrix} 0 & 0 \\ 0 & 1 \\ 0 & 0 \end{bmatrix} \end{bmatrix} \quad (A14)$$

## VII. APPENDIX B

Distance vectors are given in Table B.I. Motor parameters, mass and *field of regard* limits of the yaw and pitch gimbals are given in Table B.II. Inertia matrices calculated from the 3D model are given in Table B.III. Mass, inertia and distance parameters are calculated by assuminng inner and outer gimbals as separate rigid bodies.

TABLE B.I
Distances of the Gimbal Platform

| *Distances (in mm)* | |
|---|---|
| $\bar{r}_{Ga/a}^{(a)} = [Ga_x \; Ga_y \; Ga_z]^T$ | $[0 \; 0 \; 57.5]^T$ |
| $\bar{r}_{Gm/m}^{(m)} = [Gm_x \; Gm_y \; Gm_z]^T$ | $[0 \; -44.5 \; 0]^T$ |
| $\bar{r}_{m/a}^{(a)} = [am_x \; am_y \; am_z]^T$ | $[0 \; 44.5 \; 57.5]^T$ |
| $\bar{r}_{a/b}^{(b)} = [ba_x \; ba_y \; ba_z]^T$ | $[31.625 \; 0 \; -57.5]^T$ |

TABLE B.II
Parameters of the Yaw and Pitch Gimbal

| *Brushless DC Motor Parameters* | *Yaw* | *Pitch* |
|---|---|---|
| $K_t \; (Nm/Amp)$ | 0.0615 | 0.036 |
| $K_b \; (V/(rad/s))$ | 0.0616 | 0.0359 |
| $R(\Omega)$ | 1.42 | 1.31 |
| $L \; (mH)$ | 0.67 | 0.48 |
| $b \; (Nm/(rad/s))$ | $2.15x10^{-5}$ | $10.45x10^{-6}$ |
| *Gimbal Platform Parameters* | | |
| $m \; (kg)$ | 0.555 | 1.138 |
| *FOR (field of regard) limits* | ±45° | ±20° |

TABLE B.III
Inertia Matrices of the Yaw and Pitch Gimbal

| | *Yaw Gimbal* |
|---|---|
| $\hat{J}_a (kgm^2)$ | $\begin{bmatrix} 0.002 & -3.089x10^{-6} & 2.505x10^{-5} \\ -3.089x10^{-6} & 0.004 & 3.17x10^{-6} \\ 2.505x10^{-5} & 3.17x10^{-6} & 0.002 \end{bmatrix}$ |
| | *Pitch Gimbal* |
| $\hat{J}_m (kgm^2)$ | $\begin{bmatrix} 0.004 & 9.157x10^{-6} & 1.418x10^{-5} \\ 9.157x10^{-6} & 0.003 & -1.355x10^{-4} \\ 1.418x10^{-5} & -1.355x10^{-4} & 0.004 \end{bmatrix}$ |

## VIII. ACKNOWLEDGMENT

This work a joint study between Bilkent University and ROKETSAN Inc. (https://www.roketsan.com.tr/en). The authors thanks ROKETSAN for providing the experimetal set-up.